\documentclass[11pt, a4paper]{article}

\title{\vspace{-2cm}\bf Quantum Zeno-like Paradox for Position Measurements: A Particle
Precisely Found in Space\\ is Nowhere to be Found in Hilbert Space}

\author{Xabier Oianguren-Asua\footnote{E-mail: \texttt{xabier.oianguren-asua@math.uni-tuebingen.de}} ~~and Roderich Tumulka\footnote{E-mail: \texttt{roderich.tumulka@uni-tuebingen.de} }\vspace{0.2cm}\\ {\small \em
Mathematics Institute, Eberhard Karls University T\"ubingen,}\\ {\small \em Auf der Morgenstelle 10, 72076 T\"ubingen, Germany.
}}
\date{\small (July 6, 2026)\vspace{-0.2cm}}

\usepackage[utf8]{inputenc}
\usepackage[top=2.3cm, bottom=2.3cm, left=2.3cm, right=2.3cm]{geometry}
\usepackage{amsmath}
\usepackage{amsthm} 
\usepackage{amssymb}\usepackage{mathrsfs}
\DeclareMathAlphabet{\mathpzc}{OT1}{pzc}{m}{it}
\usepackage{enumerate}
\usepackage{dsfont} 
\usepackage{physics}

\usepackage[backend=biber, style=ieee]{biblatex}
\newcommand{\atright}[1]{%
  \unskip 
  \hfil 
  \makebox[0pt][r]{#1}
  \hfilneg 
  \hspace{0pt}
  \par 
}

\usepackage[%
framemethod=tikz,
linewidth=0.01cm,
middlelinecolor= gray,
middlelinewidth=0.02cm,
roundcorner=1cm,
topline = false,
rightline = false,
bottomline = false,
skipabove=0pt,
skipbelow=0pt,
usetwoside=false,
leftmargin=-0.6cm,
innerleftmargin=0.6cm,
rightmargin=-0.3cm,
linecolor=gray,%
backgroundcolor=none
]{mdframed}
\newmdenv[tikzsetting={draw=gray,fill=red,fill opacity=0.5}]{myenvironment}

\theoremstyle{definition}
\newtheorem{definition}{Definition}
\newtheorem{lemma}{Lemma}
\newtheorem{theorem}{Theorem}

\renewcommand\qedsymbol{ $\boldsymbol{o.\varepsilon.\delta.}$}

\newenvironment{cproof}[1][\unskip]{\begin{mdframed}
\noindent {\em Proof #1: }}{\vspace{-0.3cm}\flushright{\qedsymbol}\end{mdframed}  }

\newcommand{\Lozenge}{\vspace{-0.7cm}\flushright{{\tiny $\blacklozenge$}}}

\usepackage[dvipsnames]{xcolor}
\definecolor{tuered}{rgb}{0.627, 0.082, 0.227}

\usepackage{hyperref}
\hypersetup{
    colorlinks=true,
    linkcolor=tuered,
    filecolor=tuered,      
    urlcolor=tuered,
    citecolor=tuered
    }
\usepackage[skip=0.2cm, indent=0.3cm]{parskip}

\begin{document}
\maketitle

\begin{abstract}
    On a quantum particle in the unit interval $[0,1]$, perform a position measurement with inaccuracy $1/n$ and then a quantum measurement of the projection $|\phi\rangle\langle\phi|$ with some arbitrary but fixed normalized $\phi$. Call the outcomes $X \in[0,1]$ and $Y \in\{0,1\}$. We show that in the limit $n\to\infty$ corresponding to perfect precision for $X$, the probability of $Y=1$ tends to 0 for every $\phi$. Since there is no density matrix, pure or mixed, which upon measurement of any $|\phi\rangle\langle\phi|$ yields outcome 1 with probability 0, our result suggests that a novel type of quantum state beyond Hilbert space is necessary to describe a quantum particle after a perfect position measurement.

    \bigskip

\end{abstract}

\section{Introduction}
\label{sec:intro}

We describe a novel kind of quantum paradox, or more precisely, a counter-intuitive consequence of quantum mechanics. It is similar to the quantum Zeno effect in that it involves a limit
of increasingly precise measurements. A key difference is that it aims at measurements of \emph{position} rather
than time. It can be summarized by saying that a particle precisely
located in space is nowhere to be found in Hilbert space. We therefore call it the
\emph{spatial quantum Zeno effect}. A suggested conclusion is that a novel type of quantum state is necessary to describe the limit, one other than a vector or a density matrix in Hilbert space. This new type of state
may be worthy of further investigation.

In its simplest form (refined version below), the paradox goes as follows. 
Consider a quantum particle confined to a 1-dimensional box $[0,1]$. (It will be convenient for the notation to exclude the right end point and write $[0,1)$, but that does not matter for the argument.) The associated Hilbert space is $\mathcal{H}:=L^2\bigl([0,1),\mathbb{C},dx\bigr)=L^2\bigl( [0,1) \bigr)$. For each fixed $n\in\mathbb{N}$, consider the evenly spaced partition of $[0,1)$ into $n$ bins of length $1/n$, namely, into 
\begin{equation}
B_j^{(n)}:=\Bigl[\frac{j-1}{n},\ \frac{j}{n} \Bigr)
\end{equation}
with $j\in\{1,\ldots,n\}$. The experimental determination of the bin where the quantum particle is located constitutes a position measurement of precision $1/n$. The associated discretized position operator is $\hat{X}^{(n)}:=\sum_{j=1}^{n}\frac{j-1/2}{n}\, \hat{P}_j^{(n)}$, where $\hat{P}_j^{(n)}$ is the projection to the $j$-th bin, acting for an arbitrary $\psi\in \mathcal{H}$ as
\begin{equation}
  \Big ( \hat{P}_j^{(n)}\psi\Big)(x)=\begin{cases}
      \psi(x) & \text{ if }\ x\in B_j^{(n)}
\\ 0 & \text{ if }\ x\notin B_j^{(n)}  \end{cases}.
\end{equation}
The particle 
is initially prepared in an arbitrary wave function $\psi\in\mathcal{H}$ with $\|\psi\|=1$; then a quantum measurement of $\hat{X}^{(n)}$ is performed; call the random outcome $X$. If $X=\frac{j-1/2}{n}$, then the wave function collapses to 
\begin{equation}
    \psi' = \frac{\hat{P}_j^{(n)}\psi}{\|\hat{P}_j^{(n)}\psi\|} \,.
\end{equation}
Immediately afterwards, a quantum measurement of the operator $\hat{Y}=|\phi\rangle\langle\phi|$ is performed, for some fixed $\phi\in \mathcal{H}$ of unit norm. Since $\hat{Y}$ has eigenvalues 1 and 0, these are the possible values of the random outcome $Y$. According to the standard quantum formalism, 
\begin{align}
    \mathbb{P}^{(n)}\Big(Y=1\Big)
    &= \sum_{j=1}^n \mathbb{P}^{(n)}\Big(Y=1 \Big| X=\frac{j-1/2}{n}\Big) \, \mathbb{P}^{(n)}\Big(X=\frac{j-1/2}{n}\Big)\label{4}\\
    &= \sum_{j=1}^n \Bigl| \langle \phi|\psi'\rangle \Bigr|^2 \, \Bigl\|\hat{P}_j^{(n)}\psi\Bigr\|^2\\
    &= \sum_{j=1}^n \Bigl|  \bigl\langle \phi \big| \hat{P}_j^{(n)}\psi \bigr\rangle \Bigr|^2 \,.\label{6}
\end{align}
We prove mathematically: 
\bigskip

\begin{theorem}\label{thm:1}
    {\it For any $\psi,\phi\in\mathcal{H}$ with $\|\psi\|=1=\|\phi\|$,}
\begin{equation}\label{result.to.be.proven.in.simple.case} 
   \lim_{n\rightarrow\infty} \mathbb{P}^{(n)}\Big(Y=1\Big)=0 \,.\vspace{-0.5cm}
\end{equation}\Lozenge
\end{theorem}

In words, as the position measurement becomes arbitrarily accurate (namely, as the bins in the measurement grid become arbitrarily small), the probability that a subsequent $\hat{Y}$-measurement on the collapsed state yields $Y=1$ becomes {\em zero} for all $\phi\in\mathcal{H}$. (That is, the probability to find a contribution to the collapsed quantum state in a given 1d subspace $\mathbb{C}\phi= \mathrm{span}\{\phi\}$ of the Hilbert space $\mathcal{H}$ becomes 0, regardless of which 1d subspace we consider.) It follows that the collapsed state after an arbitrarily accurate position measurement cannot be described using a vector $\Psi$ or a density matrix $\hat{\rho}$ in $\mathcal{H}$ ---for them, there always exists a $\phi\in \mathcal{H}$ contradicting our result, namely, such that the quantum measurement of $\hat{Y}=|\phi\rangle\langle\phi|$ on them has {\em non-zero} probability of outcome 1. (Indeed, just take $\phi=\Psi$ or $\phi$ an eigenvector of $\hat\rho$ with non-zero eigenvalue.) That is the paradox.

\section{Outline of the Proof}

The mathematical proof of Theorem~\ref{thm:1} is given in the Appendix. Here is a summary of the intuitive reasons behind it. Suppose first that both $\phi$ and $\psi$ are \emph{continuous} functions. Then for large $n$,
\begin{align}
    \bigl\langle \phi \big|\hat{P}_j^{(n)}\psi \bigr\rangle 
    &= \int\limits_{(j-1)/n}^{j/n} \hspace{-2mm} \phi^*(x) \, \psi(x) \, dx\\
    &\approx \frac{1}{n}\phi^*\Bigl(\frac{j}{n}\Bigr) \, \psi\Bigl(\frac{j}{n}\Bigr) \,.
\end{align}
Taking the absolute square and summing over $j$ yields
\begin{align}
    \sum_{j=1}^n \Bigl|  \bigl\langle \phi \big| \hat{P}_j^{(n)}\psi \bigr\rangle \Bigr|^2
    &\approx \sum_{j=1}^n \frac{1}{n^2} \Bigl|\phi\Bigl(\frac{j}{n}\Bigr)\Bigr|^2 \, \Bigl|\psi\Bigl(\frac{j}{n}\Bigr)\Bigr|^2 \\
    &\approx \frac{1}{n} \int_0^1 |\phi(x)|^2 \, |\psi(x)|^2 \, dx
\end{align}
because the last sum is a Riemann sum times $1/n$. Letting $n\to\infty$ yields 0. In the rigorous proof, a variant of this reasoning is carried out for \emph{bounded} (instead of continuous) functions, and a further part of the proof shows that if this is so for bounded $\phi,\psi$, then also for arbitrary $\phi,\psi\in\mathcal{H}$.

\section{Discussion}

\paragraph{Measurement of continuous observables.}
Naively, a precise position measurement (i.e., with $n=\infty$) should yield a quantum state concentrated at a single point $x_0\in[0,1)$, and since $\mathcal{H}=L^2\bigl([0,1)\bigr)$ does not contain any such state, it is perhaps not surprising that the state of the particle we are considering does not behave like any vector in $\mathcal{H}$. On the other hand, one commonly uses wave functions outside of $\mathcal{H}$, such as the position eigenstates $|x_0\rangle$ that, when written as a wave function, would be $\delta(x-x_0)$. However, the naive collapsed state after a precise position measurement would be $\sqrt{\delta(x-x_0)}$ instead because its absolute square should have integral $=1$. The inner product of such a state with $\phi\in\mathcal{H}$ should vanish, and Theorem~\ref{thm:1} provides a precise version of this consideration. The upshot is that continuous observables (such as the position observable $\hat{X}$) not only fail to possess collapsed wave functions in $\mathcal{H}$, but they also have paradoxical joint distributions for iterated quantum measurements (such as first $\hat{X}$, then $\hat{Y}$).

\paragraph{Comparison to the quantum Zeno effect.}
In the quantum Zeno effect (e.g., \cite{Misra77}, \S 5.2.2 in \cite{tumulka}), one wants to model a measurement of the time at which a quantum particle, whose wave function is initially concentrated in some spatial region $\Omega$, first enters a given region $\Omega'$ disjoint from $\Omega$; to this end, one periodically checks whether the particle is already in $\Omega'$. The ``paradox" is that, as the frequency of observation becomes arbitrarily fast, the probability to find the particle in $\Omega'$ becomes 0 for all waiting times. Commonalities with our case are that both setups consider grids (along either the $t$ or the $x$ axis) for the purpose of discretization, both consider the continuum limit in which the mesh width of the grid tends to 0, and both lead to zero probability of a relevant type of event (detecting the particle at $t$, or ``at $\phi$'') for any value of the parameter ($t$ or $\phi$). The obvious difference is that we aim at \emph{position} rather than \emph{time} measurements, and correspondingly discretize the position rather than time axis.

\paragraph{New type of quantum state.}
For the purpose of idealization, it may be desirable to regard a theoretical limiting case of an experimental procedure also as an experimental procedure, and a limiting case of quantum states also as a quantum state. In this spirit, we may introduce a new type of quantum state as the state after the measurement of $\hat{X}^{(n)}$ in the limit $n\to\infty$, much like the Dirac $\delta$ function may be defined as a limiting case of a Gaussian function in which the width tends to 0. It is presumably well defined for this new type of quantum state which probability distribution the outcome of any subsequent quantum measurement will have, as we have seen that this is so for $\hat{Y}=|\phi\rangle \langle\phi|$: after all, in the limit $n\to\infty$, $\mathbb{P}^{(n)}(Y=0)=1$ and $\mathbb{P}^{(n)}(Y=1)=0$. Moreover, even the joint distribution of $X$ and $Y$ exists: it has all its weight in $Y=0$ with $|\psi|^2$-distribution along $X\in[0,1)$. As we have seen, such a new type of quantum state cannot be described by either a vector in Hilbert space or a density matrix. In this regard we note that the approach of rigged Hilbert spaces \cite{rigged_original, rigged_Gadella} provides an extension of Hilbert space where distributions make objects like $\delta(x-x_0)$ rigorous vectors. However, this does not seem to be the right kind of extension for our purposes because, as discussed above, we are looking for something like $\sqrt{\delta(x-x_0)}$ (rather than $\delta(x-x_0)$), which turns out to be zero as a distribution.\footnote{ Let $g\in L^1(\mathbb{R}^d)$ with $\norm{g}_{L_1}=1$ (e.g., a normalized Gaussian). For $n\in\mathbb{N}$ denote $g_n(x):=n^dg(nx)$ (e.g., the Gaussian squeezed by a factor of $n$). Then, one defines the Dirac $\delta$ as the distributional limit of $g_n$ because for all continuous and bounded function $f$, $\lim_{n\rightarrow\infty}\langle g_n,f\rangle =f(0) $. Consequently, in the distributional sense, $\sqrt{\delta}$ should be the distributional limit of $\sqrt{g_n}$, but one can check that this limit is the 0 distribution. } Instead, our limit state might be expressible as a functional on an operator algebra that yields the expectation of each observable ---after all, $\sqrt{\delta(x-x_0)}$ can be expressed \cite{Bur, Wer} as a functional on the CCR algebra (which is a proper sub-algebra of
the algebra of all bounded operators on $\mathcal{H}$ that is dense
in the strong operator topology). We leave this inquiry for future research.

\section{More General Version}

Theorem~\ref{thm:1} remains true when (i) we replace $\psi$ by a mixed state $\hat\rho$, (ii) we take grids with uneven spacing, (iii) we replace $[0,1)$ by a box $Q:=[0,1)^d$ of arbitrary dimension $d\in\mathbb{N}$ (and Hilbert space $L^2(Q)$) and (iv) we replace such a box $Q$ by the entire $\mathbb{R}^d$ (of Hilbert space $L^2(\mathbb{R}^d)$). Theorem~\ref{main.result} below shows that even after all these generalizations, the spatial quantum Zeno effect, and hence all the conclusions we drew from it, still hold.

In order to formulate the general result, we start by formalizing arbitrary rectangular grids for discretized position measurements of $d$ degrees of freedom.\vspace{0.3cm}

\begin{definition} \label{def.partition}
We define a {\em grid scheme for the unit cube $Q=[0,1)^d$} to be the choice, for each $n\in\mathbb{N}$, of a partition of $Q$ into $N_n\in\mathbb{N}$ bins $\{B_j^{(n)}\}_{j=1}^{N_n}$ such that
\begin{enumerate}[\bf (i)]
    \item  $B_j^{(n)}=\prod_{k=1}^dI_{j,k}^{(n)}$ for some intervals $I_{j,k}^{(n)}\subseteq [0,1)$ (i.e., $B_j^{(n)}$ are ``rectangular"),
    \item $\displaystyle \frac{1}{Cn} \leq \mathrm{length}(I_{j,k}^{(n)}) \leq \frac{1}{n}$ 
    for some $C>1$ independent of $n$ (i.e., the edges of the rectangles $B_j^{(n)}$ have lengths of order $1/n$).
\end{enumerate} 
A {\em grid scheme for $\mathbb{R}^d$} is defined almost verbatim---see the Appendix.\Lozenge
\end{definition}

In words, a grid scheme is a choice per $n\in\mathbb{N}$ of a rectangular grid over $Q$ or $\mathbb{R}^d$, constituted by bins of edge-lengths in the order of $1/n$. Note that as $n\rightarrow \infty$ such grids become arbitrarily fine: using $|B|$ to denote the volume of $B\subset\mathbb{R}^d$, since the length of each $I_{j,k}^{(n)}$ is at most $\frac{1}{n}$,
\begin{equation}\label{vols.of.boxes.to.zero}
|B_j^{(n)}|\leq \frac{1}{n^d}\,,
\end{equation}
 which implies that $
\lim_{n\to\infty} \max_{j} |B_j^{(n)}|= 0\,.$
Note also that even for $Q$, the number of bins $N_n$ diverges as $n\rightarrow\infty$: after all, by Eq.~\eqref{vols.of.boxes.to.zero}, there can be no fewer than $|Q|/(\frac{1}{n^d})=n^d$ bins in the $n$-th grid, i.e., $N_n\geq n^d$ (which diverges as $n\rightarrow\infty$). The simplest example of a grid scheme for $Q$ is the even subdivision into $N_n=n^d$ boxes of equal volume $1/n^d$ and edges of length $1/n$ (this, in the case $d=1$, is the grid considered in Section~\ref{sec:intro}).

For each grid scheme $\{B_j^{(n)}\}_{j=1}^{N_n}$ and each fixed $n\in \mathbb{N}$, the measurement that determines the spatial bin $B_j^{(n)}$ where the system lays, is exactly a joint position measurement (a configuration measurement) with precision of order $1/n$. It is a joint quantum measurement of the projection to the $j$-th bin for each $j$, i.e., of the multiplication operator $\hat{P}_j^{(n)}$ acting on an arbitrary wave function $\psi$ as
\begin{equation}
  \Big ( \hat{P}_j^{(n)}\psi\Big)(x)=\mathds{1}_{B_j^{(n)}}(x) \: \psi(x),
\end{equation}
where we denote by $\mathds{1}_{B}$ the indicator function of the set $B\subseteq \mathbb{R}^d.$ 
Equivalently, this is a joint quantum measurement of the discretized position operators $\hat{X}_k^{(n)}:=\sum_{j=1}^{N_n}x_{j,k}^{(n)}\, \hat{P}_j^{(n)}$, where $x_{j,k}^{(n)}$ can be taken to be any point of $I_{j,k}^{(n)}$ and $k=1,\ldots,d$ labels the dimension. 

Now, let us consider an arbitrary mixed state $\hat{\rho}$ as the initial quantum state (either on $L^2(Q)$ or $L^2(\mathbb{R}^n)$). After a joint position measurement of $\hat{X}_1^{(n)},....,\hat{X}_d^{(n)}$ on $\hat{\rho}$, we measure $\hat{Y}=|\phi\rangle\langle\phi|$ on the collapsed state and call the outcome $Y$, which can be 0 or 1. By a calculation analogous to Eqs.~\eqref{4}--\eqref{6}, one gets the standard prediction
\begin{equation}
    \label{prob.Y.1.general.case}
    \mathbb{P}^{(n)}\Big(Y=1\Big)=\Bigl\langle \phi \:\Big|\: \sum_{j=1}^{N_n}\hat{P}_j^{(n)}\hat{\rho} \hat{P}_j^{(n)} \: \Big| \: \phi \Bigr\rangle.
\end{equation}

\begin{theorem}\label{main.result}
    {\it For any density matrix  $\hat{\rho}$ on $L^2(Q)$, any unit vector $\phi\in L ^2(Q)$, and any grid scheme $\{B_j^{(n)}\}_{j=1}^{N_n}$ for $Q$}:\vspace{-0.1cm}
    \begin{equation}
\lim_{n\rightarrow\infty}\mathbb{P}^{(n)}\Big(Y=1\Big)=0.
    \end{equation}
{\it The same statement holds if we replace $Q$ by $\mathbb{R}^d$.}\Lozenge
\end{theorem}

\section{Conclusions}

If the accuracy of a position measurement for a quantum particle becomes arbitrarily good, any subsequent 1d ``yes-no" projective measurement yields ``no" with certainty. Such ``yes-no" measurements are projections to subspaces of Hilbert space, so this implies that a perfectly localized particle in physical space is nowhere to be found in Hilbert space ---not even as a density matrix. Consequently, a novel type of quantum state is necessary to describe the limiting experimental setting (presumably a functional on the relevant algebra of operators), and this deserves further investigation.

\section*{Appendix: Mathematical Proofs}

Since Theorem~\ref{thm:1} is a special case of Theorem~\ref{main.result}, it suffices to prove the latter. 
It is well known that because $Q$ has finite volume, $L^\infty(Q) \subseteq L^2(Q)\subseteq L^1(Q)$.\vspace{0.3cm}
\begin{definition}
    Given $f\in L^1(Q)$ and a grid scheme  $\{B_j^{(n)}\}_{j=1}^{N_n}$ for $Q$, we define the ``discretization of $f$ by the $n$-th grid" to be the function $f_n:Q\rightarrow\mathbb{C}$,
    \begin{equation}
        f_n(x):=\sum_{j=1}^{N_n}\underbrace{\frac{1}{|B_j^{(n)}|}\int_{y\in B_j^{(n)}} f(y) \,dy}_{\text{average of $f$ in $B_j^{(n)}$}}\;\mathds{1}_{B_j^{(n)}}(x)\quad \quad \text{ for } x\in Q.\vspace{-0.2cm}
    \end{equation}(Informally, it is the ``bar chart" taking at each point of $B_j^{(n)}$ the average of $f$ in $B_j^{(n)}$ as value.)\Lozenge
\end{definition}\vspace{0.3cm}
\begin{lemma}\label{leb.diff.lemma}
    {\it Let there be an arbitrary $f\in L^\infty(Q)$ and let $\{B_j^{(n)}\}_{j=1}^{N_n}$ be an arbitrary grid scheme for $Q$. Then, $f_n\in L^{{\infty}}(Q)$ and $\norm{f_n-f}_{L^2}\xrightarrow[n\rightarrow\infty]{}0$.}\Lozenge\vspace{0.3cm}
\end{lemma}

\begin{cproof}
First, note that for each $x\in Q$ and $n\in \mathbb{N}$, because $\{B_j^{(n)}\}_{j=1}^{N_n}$ is a partition of $Q$, there exists a unique $j_{x}^n\in \{1,...,N_n\}$ such that $x\in B_{j_x^n}^{(n)}$. 

Next, for each $x\in Q$,\vspace{-0.3cm}
\begin{align}
 | f_n(x)|&=\Bigg|\sum_{j=1}^{N_n}\frac{\mathds{1}_{B_j^{(n)}}(x)}{|B_j^{(n)}|}\int_{y\in B_j^{(n)}} f(y) \, dy\Bigg|
 \end{align}
\begin{align}
 &\leq \sum_{j=1}^{N_n}\frac{\mathds{1}_{B_j^{(n)}}(x)}{|B_j^{(n)}|}\int_{y\in B_j^{(n)}} |f(y)| \, dy\\
 &\stackrel{\ref{only.one.survives}}{=}\frac{1}{|B_{j_x^n}^{(n)}|}\int_{y\in B_{j_x^n}^{(n)}} |f(y)| \, dy\\
 &\leq \norm{f}_{L ^\infty},
\end{align}
 \stepcounter{mpfootnote}  \footnotetext{\label{only.one.survives}By definition, $\mathds{1}_{B_j^{(n)}}(x)=0$ for all $j\in \{1,...,N_n\}$ except for $j=j_x^n$, where $\mathds{1}_{B_{j_x^n}^{(n)}}(x)=1$. }
and hence $\norm{f_n}_{L^\infty}\leq \norm{f}_{L^\infty}$, such that $f_n\in L^\infty(Q)$, and a fortiori $f_n\in L^2(Q)$.

Now recall that for each $x\in Q$, $B_{j_x^n}^{(n)}=\prod_{k=1}^nI_{j_x^n,k}^{(n)}$ for some intervals $I_{j_x^n,k}^{(n)}\subseteq[0,1)$ of length in $[\frac{1}{Cn}, \,\frac{1}{n}]$. This implies that the longest diagonal of $B_{j_x^n}^{(n)}$ has at most length $\frac{\sqrt{d} }{n}$ and thus that $B_{j_x^n}^{(n)}$ is contained in the open ball of radius $R_n:=\frac{2\sqrt{d} }{n}$ around $x$, which we denote by $V_{R_n}(x)$. Moreover, for the $n$-independent number $K:=\frac{1}{C^d(2\sqrt{d} )^d|V_{1}(0)|}$, 
  \begin{equation}
   |B_{j_x^n}^{(n)}|\stackrel{\footnote{By definition, the edges of $B_{j_x^n}^{(n)}$ have a length no smaller than $1/(Cn)$, so the volume of $B_{j_x^n}^{(n)}$ is at least $\frac{1}{(Cn)^d}$.  }}{\geq}  \frac{1}{(Cn)^d}= K\Big(\frac{2\sqrt{d} }{n}\Big)^d|V_{1}(0)|= K (R_n)^d|V_1(0)|=K|V_{R_n}(x)|.
  \end{equation}
This condition, together with the fact that $f$ is integrable\footnote{By assumption, $f\in L^\infty(Q)$, which implies that $f\in L^1(Q)$.} ---which by Theorem 1.6.19 in Ref.~\cite{tao} implies that almost every (a.$\;$e.) $x\in Q$ is a Lebesgue point--- allow us to apply Ex.$\;$1.6.15 in Ref.~\cite{tao}, which is a corollary of the Lebesgue differentiation theorem. This corollary proves that for a.$\;$e.$\;x\in Q$,
\begin{equation}\label{Lebsegue.diff.thm}
    \lim_{n\rightarrow\infty}\Bigg(\frac{1}{|B_{j^n_x}^{(n)}|}\int_{y\in B_{j_x^{n}}^{(n)}}f(y) \, dy\Bigg)=f(x).
\end{equation}

Now, note the following.\begin{enumerate}
    \item[(i)] Point-wise, for a.$\;$e.$\;x\in Q$,\vspace{-0.1cm}
\begin{align}
|f_n(x)-f(x)|&=\Bigg|\sum_{j=1}^{N_n}\frac{\mathds{1}_{B_j^{(n)}}(x)}{|B_j^{(n)}|}\int_{y\in B_j^{(n)}}f(y)\, dy -f(x)\Bigg|\\
&\stackrel{\ref{only.one.survives}}{=}\Bigg|\frac{1}{|B_{j_x^n}^{(n)}|}\int_{y\in B_{j_x^n}^{(n)}}f(y)\, dy -f(x)\Bigg|\xrightarrow[n\rightarrow\infty]{\eqref{Lebsegue.diff.thm}}0.
\end{align}
\item[(ii)] The constant (and hence $L^\infty)$ function $g(x)\equiv 2\norm{f}_{L^\infty}$ satisfies that for a.$\;$e.$\;x\in Q$, 
\begin{equation}
|f_n(x)-f(x)|\leq \frac{1}{|B_{j_x^n}^{(n)}|}\int_{y\in B_{j_x^n}^{(n)}}\Big|f(y)\Big| \, dy +\Big|f(x)\Big|\leq 2\norm{f}_{L^\infty}\:.
\end{equation}
\end{enumerate}  By virtue of (i) and (ii), $g$ is a dominating function in $L^2(Q)$ that grants the application of the dominated convergence theorem in the last step of the following chain of equations:
\begin{equation}
\lim_{n\rightarrow\infty}\norm{f_n-f}_{L^2}^{2}=\lim_{n\rightarrow\infty}\int_{x\in Q}|f_n(x)-f(x)|^{2} \, dx=0.\vspace{-0.7cm}
\end{equation}
\end{cproof}
\vspace{0.3cm}

We are now ready to prove the result for $Q$ in the special case that $\hat\rho=|\psi\rangle\langle\psi|$ is pure and both $\psi$ and $\phi$ are (essentially) bounded functions rather than general $L^2$ functions.\vspace{0.3cm}

\begin{lemma}\label{result.for.Linfty}
       {\it Given arbitrary $\psi,\phi\in L ^\infty(Q)$ and an arbitrary grid scheme $\{B_j^{(n)}\}_{j=1}^{N_n}$ for $Q$:} \vspace{-0.1cm}
    \begin{equation}
        \lim_{n\rightarrow\infty}\sum_{j=1}^{N_n}\Big|\langle \phi| \hat{P}_j^{(n)}\psi\rangle\Big|^2=0.\vspace{-0.3cm}
    \end{equation}\Lozenge\vspace{0.3cm}
\end{lemma}

\begin{cproof}
First, define $f(x):=\overline{\phi(x)}\psi(x)$ for $x\in Q$. Then $f\in L ^\infty(Q)$ because for a.$\,$e.$\;x\in Q$, $|f(x)|=|\phi(x)| \, |\psi(x)|\leq \|\phi\|_{L^\infty} \|\psi\|_{L^\infty}$.
Hence, $f\in L^2(Q)$ and moreover, we can apply Lemma~\ref{leb.diff.lemma} to $f$ to get that $\norm{f_n-f}_{L^2}\xrightarrow[n\rightarrow
\infty]{}0$. This implies by the reverse triangle inequality that $\norm{f_n}_{L^2}\rightarrow\norm{f}_{L^2}$ and thus $\norm{f_n}_{L^2}^2\rightarrow\norm{f}_{L^2}^2.$

Now,
\begin{align}
\norm{f_n}_{L^2}^2&=\int_{x\in Q}\Bigg| \sum_{j=1}^{N_n}\frac{\mathds{1}_{B_j^{(n)}}(x)}{|B_j^{(n)}|}\underbrace{\int_{y\in B_j^{(n)}} f(y) \, dy}_{\int_{y\in B^{(n)}_j}\overline{\phi(y)} \, \psi(y) \, dy}\;\Bigg|^2dx\\
&=\int_{x\in Q}\Bigg| \sum_{j=1}^{N_n}\frac{\mathds{1}_{B_j^{(n)}}(x)}{|B_j^{(n)}|}\langle \phi| \hat{P}_j^{(n)}\psi\rangle\;\Bigg|^2dx\\
\intertext{[use that $\{\mathds{1}_{B_j^{(n)}}\}_{j=1}^{N_n}$ have disjoint supports]}
&= \sum_{j=1}^{N_n}\int_{x\in Q}\Bigg| \frac{\mathds{1}_{B_j^{(n)}}(x)}{|B_j^{(n)}|}\langle \phi| \hat{P}_j^{(n)}\psi\rangle\;\Bigg|^2dx \label{stepstar}\\
&=\sum_{j=1}^{N_n}\frac{|B_j^{(n)}|}{|B_j^{(n)}|^2} \bigl|\langle\phi| \hat{P}^{(n)}_j\psi\rangle \bigr|^2\\
&=\sum_{j=1}^{N_n}\frac{1}{|B_j^{(n)}|} \bigl|\langle\phi| \hat{P}^{(n)}_j\psi\rangle \bigr|^2.
\end{align}
But then $\lim_{n\rightarrow\infty}\norm{f_n}_{L^2}^2=\norm{f}^2_{L^2}$ means that the sequence $\Big(\sum_{j=1}^{N_n} \frac{1}{|B_j^{(n)}|}\Big|\langle \phi| \hat{P}_j^{(n)}\psi\rangle\Big|^2\Big)_{n\in \mathbb{N}}$ converges to the finite number $\norm{f}_{L^2}^2$. Using this result to obtain Eq.~\eqref{stepstarstar} below, together with the fact that $(1/n^d)_{n\in \mathbb{N}}$ converges to 0 and the continuity of the product of complex numbers,
\begin{align}
\sum_{j=1}^{N_n}\Big|\langle \phi| \hat{P}_j^{(n)}\psi\rangle\Big|^2
&=\sum_{j=1}^{N_n}\frac{|B_j^{(n)}|}{|B_j^{(n)}|}\Big|\langle \phi| \hat{P}_j^{(n)}\psi\rangle\Big|^2\\
&\stackrel{\eqref{vols.of.boxes.to.zero}}{\leq} \frac{1}{n^d}\ \sum_{j=1}^{N_n} \frac{1}{|B_j^{(n)}|}\Big|\langle \phi| \hat{P}_j^{(n)}\psi\rangle\Big|^2\\
&\xrightarrow[n\rightarrow\infty]{}0\cdot\norm{f}_{L^2}^2=0. \label{stepstarstar}
\end{align}\vspace{-0.8cm}
\end{cproof}\vspace{0.3cm}

With that, we can prove the spatial quantum Zeno effect for any pure state $\hat{\rho}=|\psi\rangle\langle \psi|$ on $L^2(Q)$.\vspace{0.3cm}

\begin{lemma}\label{main.result.in.Q.for.pure.states}
    {\it Given arbitrary $\psi,\phi\in L ^2(Q)$ and an arbitrary grid scheme $\{B_j^{(n)}\}_{j=1}^{N_n}$ for $Q$},\vspace{-0.1cm}
    \begin{equation}
\lim_{n\rightarrow\infty}\sum_{j=1}^{N_n}\Big|\langle \phi| \hat{P}_j^{(n)}\psi\rangle\Big|^2=0.\vspace{-0.4cm}
    \end{equation}\Lozenge
\end{lemma}

\begin{cproof} It is well known that the smooth compactly supported functions $\mathcal{C}^\infty_0(Q)$ are dense in $L^2(Q)$ (see Prop.$\;$8.17 in Ref.~\cite{folland} for the case of $\mathbb{R}^d$ ---with the right identifications, one can restrict this result to $Q$). In particular, all functions on $\mathcal{C}_0^\infty(Q)$ are continuous maps with compact support and thus, they are bounded functions. As such, $\mathcal{C}_0^\infty(Q)\subseteq L^\infty(Q)$, proving that $L^\infty(Q)$ is also dense in $L^2(Q).$

By this denseness, for an arbitrary $\varepsilon>0$, one can successively find $\psi_\varepsilon,\phi_\varepsilon \in L^\infty(Q)$ such that $\norm{\psi-\psi_\varepsilon}^2\leq \frac{\varepsilon}{6\norm{\phi}^2}$ and $\norm{\phi-\phi_\varepsilon}^2\leq \frac{\varepsilon}{12\norm{\psi_\varepsilon}^2}.$ In particular, by Lemma \ref{result.for.Linfty}, $\sum_{j=1}^{N_n} \frac{1}{|B_j^{(n)}|}\Big|\langle \phi_\varepsilon| \hat{P}_j^{(n)}\psi_\varepsilon\rangle\Big|^2$ $\xrightarrow[n\rightarrow
\infty]{}0$, so there is $n_\varepsilon\in \mathbb{N}$ such that $\sum_{j=1}^{N_n} \frac{1}{|B_j^{(n)}|}\Big|\langle \phi_\varepsilon| \hat{P}_j^{(n)}\psi_\varepsilon \rangle\Big|^2\leq \frac{\varepsilon}{12 }$ for all $n\geq n_\varepsilon.$ Hence, for all $n\geq n_\varepsilon$,
\begin{align}
\sum_{j=1}^{N_n}\Big|\langle & \phi| \hat{P}^{(n)}_j\psi\rangle\Big|^2\ \stackrel{\ref{triangl.ineq.sqr}}{\leq}\  \sum_{j=1}^{N_n}2\Big(\Big|\langle \phi| \hat{P}^{(n)}_j(\psi-\psi_\varepsilon)\rangle\Big|^2+\Big|\langle \phi| \hat{P}^{(n)}_j\psi_\varepsilon\rangle\Big|^2\Big)\\
&\stackrel{\ref{triangl.ineq.sqr}}{\leq} \sum_{j=1}^{N_n}2\Big(\Big|\langle \hat{P} ^{(n)}_j\phi| \psi-\psi_\varepsilon\rangle\Big|^2+2\Big|\langle \phi-\phi_\varepsilon| \hat{P}^{(n)}_j\psi_\varepsilon\rangle\Big|^2+2\Big|\langle \phi_\varepsilon| \hat{P}^{(n)}_j\psi_\varepsilon\rangle\Big|^2\Big)\\
&\leq 2\underbrace{\sum_{j=1}^{N_n}\norm{\hat{P}^{(n)}_j\phi}^2}_{=\norm{\phi}^2\text{ by \ref{id.resolution}}}\norm{\psi-\psi_\varepsilon}^2+4\norm{\phi-\phi_\varepsilon}^2\underbrace{\sum_{j=1}^{N_n}  \norm{\hat{P}^{(n)}_j\psi_\varepsilon}^2}_{=\norm{\psi_\varepsilon}^2\text{ by \ref{id.resolution}}}+4\sum_{j=1}^{N_n}\frac{|B_j^{(n)}|}{|B_j^{(n)}|}\Big|\langle \phi_\varepsilon| \hat{P}^{(n)}_j\psi_\varepsilon\rangle\Big|^2\\
\intertext{[now use the definition of $\psi_\varepsilon,\phi_\varepsilon$ and Eq.~\eqref{vols.of.boxes.to.zero}]}
&\leq\ \frac{\varepsilon}{3}+\frac{\varepsilon}{3}+ 4\;\frac{1}{n^d}\sum_{j=1}^{N_n}\frac{1}{|B_j^{(n)}|}\Big|\langle \phi_\varepsilon| \hat{P}^{(n)}_j\psi_\varepsilon\rangle\Big|^2\ \stackrel{(n\geq n_\varepsilon)}{\leq}\ \frac{2\varepsilon}{3}+\frac{1}{n^d}\frac{\varepsilon}{3}\ \leq \ \varepsilon.
\end{align}
\stepcounter{mpfootnote}
\footnotetext{\label{triangl.ineq.sqr} Successively, we add and subtract $\langle \phi| \hat{P}^{(n)}\psi_\varepsilon\rangle$ and then $\langle \phi_\varepsilon| \hat{P}^{(n)}\psi_\varepsilon\rangle$, using after each such operation that for all $a,b\in\mathbb{C}$, $|a-b|^2=|a|^2+|b|^2-2\mathrm{Re}\, \overline{a}b\leq |a|^2+|b|^2+2|a||b|\leq 2|a|^2+2|b|^2$. }
\stepcounter{mpfootnote}
\footnotetext{\label{id.resolution}Let $\eta\in L^2(Q)$ be arbitrary. Then, using that $\hat{P}_j^{(n)}$ are self-adjoint projectors and the obvious property $\hat{I}=\sum_{j=1}^{N_n}\hat{P}_j^{(n)}$, we get that $\sum_{j=1}^{N_n}\norm{\hat{P}_j^{(n)}\eta}^2=\langle \eta| \sum_{j=1}^{N_n}\hat{P}_j^{(n)}\eta\rangle=\langle \eta|\eta\rangle=\norm{\eta}^2.$}
But by definition, this is to say that $\displaystyle \sum_{j=1}^{N_n}\Big|\langle \phi| \hat{P} ^{(n)}_j\psi\rangle\Big|^2\xrightarrow[n\rightarrow\infty]{}0$.\vspace{-0.5cm}
\end{cproof}
\vspace{0.3cm}
The generalization of Lemma \ref{main.result.in.Q.for.pure.states} to translated and scaled cubes (namely, to arbitrary rectangles in $\mathbb{R}^d$) is straightforward. 

Next, we are going to split $\mathbb{R}^d$ into countably many cubes and equip each of them with a grid scheme, making together a family of ``position measurement grids" for the whole $\mathbb{R}^d$.
\vspace{0.3cm}
\begin{definition}\label{def.partition.Rd}
    Given a translated unit cube $\widetilde{Q}:=\prod_{k=1}^d[a_k,\ a_k+1)$ for some $a_k\in\mathbb{R}$, we define a {\em grid scheme for $\widetilde{Q}$} to be the set of translations $\widetilde{V}_{j}^{(n)}:={V}_{j}^{(n)}+(a_1,...,a_d)$ of an arbitrary grid scheme for $[0,1)^d$, $\{{V}_{j}^{(n)}\}_{j=1}^{N_n}$. 
    
   \noindent Then, we define a {\em grid scheme for $\mathbb{R}^d$} to be a partition of $\mathbb{R}^d$ into translated unit cubes $\{Q_\ell\}_{\ell\in \mathbb{N}}$ and a choice of {\em grid scheme} per $Q_\ell$, say, $\{\widetilde{V}_{\ell,j}^{(n)}\}_{\ell\in\mathbb{N},\ j\in\{1,...,N_\ell^{(n)}\}}$ $(n\in\mathbb{N})$. Since for each $n\in\mathbb{N}$, $\{V_{\ell,j}^{(n)}\}_{\ell\in\mathbb{N},\ j\in\{1,...,N_\ell^{(n)}\}}$ is a countable set, we will write $\{V_{\ell,j}^{(n)}\}_{\ell\in\mathbb{N},\ j\in\{1,...,N_\ell^{(n)}\}}=\{B_j^{(n)}\}_{j\in\mathbb{N}}$ and define $J_\ell^{(n)}:=\{j\in \mathbb{N}\ |\ B_j^{(n)}\subseteq Q_\ell\}$ (such that $|J_\ell^{(n)}|=N_\ell^{(n)}$).\Lozenge\vspace{0.1cm}
\end{definition}
For each fixed $n\in\mathbb{N}$, $\{B_j^{(n)}\}_{j\in\mathbb{N}}$ is trivially a partition of $\mathbb{R}^d$ into rectangles of edges in the order of $1/n$. The only restriction we put by the inductive definition is that the interior of no $B_j^{(n)}$ is allowed to intersect the boundary of a cube $Q_\ell$. Although this made the last definition a bit cumbersome, we imposed it to make the following statements and proofs notationally simpler.
 \vspace{0.3cm}
\begin{lemma}\label{general.result.in.Rd.for.pure}

    {\em Given arbitrary $\psi,\phi\in L^2(\mathbb{R}^d)$ and an arbitrary grid scheme $\{B_{j}^{(n)}\}_{j\in \mathbb{N}}$ for $\mathbb{R}^d$}:
    \begin{equation}
        \lim_{n\rightarrow\infty}\sum_{j=1}^\infty\Big|\langle \phi|\hat{P}_j^{(n)}\psi\rangle \Big|^2=0.\vspace{-0.4cm}
    \end{equation}\Lozenge\vspace{0.3cm}
\end{lemma}
\begin{cproof}
First, note the following observation. Given a measurable infinite set $A\subseteq\mathbb{R}^d$, a partition $\{A_M\}_{M\in\mathbb{N}}$ for $A$ and a non-negative integrable function $f:\mathbb{R}^d\rightarrow[0,+\infty]$,
\begin{enumerate}
    \item[(i)] for all $x\in A$, $\sum_{j=1}^M\mathds{1}_{A_j}(x) f(x)$ equals $\mathds{1}_A(x) f(x)$ from some $M$ onward,\footnote{Namely, for all $M$ if $x\not\in A$ and otherwise, after the $M$ such that $x\in A_M$.}
    \item[(ii)] $\sum_{j=1}^M\mathds{1}_{A_j} f$ is a monotonously increasing sequence of measurable functions. 
\end{enumerate}
Consequently, the monotone convergence theorem implies that: 
\begin{equation}\label{claim}
\lim_{M\rightarrow\infty}  \sum_{j=1}^M  \Big\|\mathds{1}_{A_n} f\Big\|_{L^1(\mathbb{R^d})}=\lim_{M\rightarrow\infty}\int_{x\in \mathbb{R}^d}\sum_{j=1}^M\mathds{1}_{A_n}(x)f(x)dx=\int_{x\in\mathbb{R}^d}\mathds{1}_A(x)f(x)dx=\Big\|\mathds{1}_A f\Big\|_{L^1(\mathbb{R}^d)}.
\end{equation}
This proves the last step of the following.
    \begin{align}
            \sum_{j=1}^M|\langle \phi| \hat{P}_j^{(n)}\psi\rangle |^2&\leq \|\phi\|^2\sum_{j=1}^M\|\hat{P}_j^{(n)}\psi\|^2\quad\\ &=\|\phi\|^2\sum_{j=1}^M\Big\|\mathds{1}_{B_j^{(n)}}|\psi|^2\Big\|_{L^1}\ \ \xrightarrow[M\rightarrow\infty]{\eqref{claim}}\ \ \|\phi\|^2\|\psi\|^2.
    \end{align}
Hence, for each $n\in\mathbb{N}$, the symbol $\sum_{j=1}^\infty\Big|\langle \phi|\hat{P}_j^{(n)}\psi\rangle \Big|^2$ in the Lemma's statement is finite and well defined. In particular, it is an absolutely convergent series, which implies that one can add its terms in finite ``chunks" before adding them together and the result will still be the same. That is, using the notation of Definition \ref{def.partition.Rd},
\begin{equation}\label{in.chunks}
    \sum_{j=1}^\infty|\langle \phi| \hat{P}_j^{(n)}\psi\rangle |^2=\sum_{\ell=1}^\infty\sum_{j\in J_\ell^{(n)}} |\langle \phi| \hat{P}_j^{(n)}\psi\rangle |^2.
\end{equation}
Now, define $h_n(\ell):=\sum_{j\in J_\ell^{(n)}} |\langle \phi| \hat{P}_j^{(n)}\psi\rangle |^2$ and note the following.
\begin{enumerate}[(i)]
    \item Point-wise, for each fixed $\ell\in \mathbb{N}$,
    \begin{equation}
        \lim_{n\rightarrow\infty}h_n(\ell)\ =\ \lim_{n\rightarrow\infty}\sum_{j\in J_\ell^{(n)}} |\langle \phi| \hat{P}_j^{(n)}\psi\rangle |^2\ \stackrel{\text{(Lemma \ref{main.result.in.Q.for.pure.states})}}{=}\ 0.
    \end{equation}
    \item For each $\ell\in\mathbb{N}$ and independently of $n\in\mathbb{N}$,
    \begin{equation}
        |h_n(\ell)|\ \leq\ \|\phi\|^2\sum_{j\in J_\ell^{(n)}}\|\hat{P}_j^{(n)}\psi\|^2\ \stackrel{\eqref{claim}}{=}\ \|\phi\|^2\Big\|\mathds{1}_{Q_\ell}|\psi|^2\Big\|_{L^1(\mathbb{R}^d)}\ =:g(\ell).
    \end{equation}
    In particular, denoting by $d\nu$ the counting measure of $\mathbb{N}$, 
    \begin{equation}
\|g\|_{L^1(\mathbb{N},d\nu)}\ =\ \|\phi\|^2\lim_{M\rightarrow\infty}\sum_{\ell=1}^M\Big\|\mathds{1}_{Q_\ell}|\psi|^2\Big\|_{L^1(\mathbb{R}^d)}\ \stackrel{\eqref{claim}}{=}\ \|\phi\|^2\|\psi\|^2\ <+\infty.
    \end{equation}
    Hence, $g$ is a dominating function in $ L^1(\mathbb{N},d\nu)$ for the sequence of functions $h_n:\mathbb{N}\rightarrow\mathbb{R}$. 
    \end{enumerate}

By the dominated convergence theorem, (i) and (ii) imply that
    \begin{equation}\label{dominated.conv.Rd}
        \lim_{n\rightarrow\infty}\int_{\ell\in \mathbb{N}}h_n(\ell)d\nu=0.\vspace{-0.2cm}
    \end{equation}
    Putting it all together,
    \begin{equation}
        \lim_{n\rightarrow\infty}   \sum_{j=1}^\infty|\langle \phi| \hat{P}_j^{(n)}\psi\rangle |^2\ \stackrel{\eqref{in.chunks}}{=}\  \lim_{n\rightarrow\infty}\sum_{\ell=1}^\infty\sum_{j\in J_\ell^{(n)}} |\langle \phi| \hat{P}_j^{(n)}\psi\rangle |^2\ \stackrel{\text{(by def.)}}{=}\ \lim_{n\rightarrow\infty}\int_{\ell\in \mathbb{N}}h_n(\ell)d\nu\ \stackrel{\eqref{dominated.conv.Rd}}{=}\ 0.\vspace{-0.5cm}
    \end{equation} 
\end{cproof}
\vspace{0.3cm}

 Finally, let us lift these results to the case of arbitrary density matrices.\vspace{0.3cm}

 \noindent {\bf Theorem \ref{main.result}.}
{\em  For any positive trace-class operator $\hat{\rho}$ on $L^2(Q)$, any vector $\phi\in L ^2(Q)$, and any grid scheme $\{B_j^{(n)}\}_{j=1}^{N_n}$ for $Q$,}\vspace{-0.2cm}
    \begin{equation}
\lim_{n\rightarrow\infty}\Big\langle \phi\Big|\sum_{j=1} ^{N_n}\hat{P}_j^{(n)}\hat{\rho} \hat{P}_j^{(n)}\phi\Big\rangle=0.
    \end{equation}
{\em The same statement holds if we replace $Q$ by $\mathbb{R}^d$ (consequently setting $N_n=+\infty$).}
\atright{\tiny $\blacklozenge$\vspace{0.3cm}}
\vspace{0.3cm}

\begin{cproof}
We give the proof for $Q$ and $\mathbb{R}^d$ together. The only difference will be whether $N_n$ is finite or infinite for every $n\in\mathbb{N}$.

By Theorems 1.1 and 1.2 in Ref.~\cite{simon.trace.ideals}, because $\hat\rho$ is a self-adjoint and trace-class (hence, compact) operator, there exists an ONB of eigenvectors $\{\psi_\ell\}_{\ell\in\mathbb{N}}\subset L^2(Q)$ (resp. in $L^2(\mathbb{R}^d)$) with associated eigenvalues $p_\ell\geq 0$, such that,
      \begin{equation}\label{tal}
          \hat{\rho}=\sum_{\ell=1}^\infty p_\ell |\psi_\ell\rangle \langle \psi_\ell|,
      \end{equation}  
     where the series is taken in the operator norm. In particular, by definition, the trace-norm of $\hat{\rho}$, $\norm{\hat\rho}_{tr}$, equals $\sum_{\ell=1}^\infty p_\ell$. Using the sequential continuity of the inner product to get Eqs.~\eqref{cual} and \eqref{pascual},\vspace{-0.3cm}
     \begin{align}
         \label{cual} \Big\langle \phi\Big|\sum_{j=1} ^{N_n}\hat{P}_j^{(n)}\hat{\rho} \hat{P}_j^{(n)}\phi\Big\rangle&=\sum_{j=1}^{N_n}\Big\langle \phi\Big|\hat{P}_j^{(n)}\hat{\rho} \hat{P}_j^{(n)}\phi\Big\rangle\\    
& \stackrel{\eqref{tal}}{=}\sum_{j=1} ^{N_n}  \sum_{\ell=1}^\infty p_\ell \Big \langle \phi\:\Big|\: \hat{P}_j^{(n)}\psi_\ell\Big\rangle\Big\langle \psi_\ell \:\Big|\:  \hat{P}_j^{(n)} \phi\Big\rangle\\
\label{pascual} & =\sum_{j=1} ^{N_n}  \sum_{\ell=1}^\infty p_\ell\ \Big|\langle \phi| \hat{P}_j^{(n)}\psi_\ell\rangle\Big|^2\\
\intertext{[considering the series as iterated integrals in the counting measure $d\nu$ of $\mathbb{N}$, since the integrand is non-negative, we can apply the Tonelli theorem (2.37 in Ref.~\cite{folland}) to switch the order of summation]}
& = \sum_{\ell=1}^\infty\sum_{j=1} ^{N_n} p_\ell\Big|\langle \phi| \hat{P}_j^{(n)}\psi_\ell\rangle\Big|^2\\
\intertext{[we define $h_n(\ell):=\sum_{j=1} ^{N_n} p_\ell\Big|\langle \phi| \hat{P}_j^{(n)}\psi_\ell\rangle\Big|^2$ and write the resulting series as an integral]}
&\label{azkena} =\int_{\ell\in\mathbb{N}}h_n(\ell)d\nu.
     \end{align}
  Finally, note that:
    \begin{enumerate}
        \item[(i)] Point-wise, for each fixed $\ell\in\mathbb{N}$,\vspace{-0.3cm}
        \begin{equation}
        \lim_{n\rightarrow\infty}h_n(\ell)\ \stackrel{\text{(by def.)}}{=}\ p_\ell\lim_{n\rightarrow\infty} \sum_{j=1}^{N_n} \Big|\langle \phi| \hat{P}_j^{(n)}\psi_\ell\rangle\Big|^2\ \stackrel{\text{(Lemmas \ref{main.result.in.Q.for.pure.states} \& \ref{general.result.in.Rd.for.pure})}}{=}\ 0.\vspace{-0.3cm}
        \end{equation}
        
        \item[(ii)] For each $\ell\in \mathbb{N}$ and independently of $n\in\mathbb{N}$,\vspace{-0.2cm}
        \begin{equation}
        |h_n(\ell)|\stackrel{\text{(Cauchy-Schwarz)}}\leq p_\ell \norm{\phi}^2\sum_{j=1}^{N_n}  \norm{\hat{P}_j^{(n)}\psi_\ell}^2\ \stackrel{\eqref{claim}}{=}\ p_\ell\norm{\phi}^2\norm{\psi_\ell}^2\ \stackrel{(\norm{\psi_\ell}=1)}{=}\ p_\ell\norm{\phi}^2.
        \end{equation}
        In particular, defining $g(\ell):=p_\ell\norm{\phi}^2$,
        \begin{equation}
            \norm{g}_{L^1(\mathbb{N},d\nu)}\ =\ \norm{\phi}^2\sum_{\ell=1}^\infty |p_\ell|\ =\ \norm{\phi}^2\norm{\hat\rho}_{tr}\ <\ +\infty.
        \end{equation}
        Hence, $g$ is a dominating function in $L^1(\mathbb{N},d\nu)$ for the sequence of functions $h_n$.
    \end{enumerate}
    By (i) and (ii), we can use the dominated convergence theorem to get the last step of the following: 
    \begin{equation}
     \lim_{n\rightarrow\infty}   \Big\langle \phi\Big|\sum_{j=1} ^{N_n}\hat{P}_j^{(n)}\hat{\rho} \hat{P}_j^{(n)}\phi\Big\rangle\ \stackrel{\eqref{azkena}}{=}\ \lim_{n\rightarrow\infty}\int_{\ell\in\mathbb{N}}h_n(\ell)d\nu\ =\ 0.\vspace{-0.5cm}
    \end{equation}
\end{cproof}

\section*{Acknowledgements}
 XOA gratefully acknowledges the support of a fellowship from the ``la Caixa"
Foundation (ID PFA25-00220F \& code LCF/BQ/PFA25/11000040).

\printbibliography

\end{document}